\documentclass{llncs}
\usepackage{graphicx,epsfig,color}
\usepackage[shortend,ruled]{algorithm2e}
\usepackage{ifthen,verbatim,array,multirow}
\usepackage{latexsym,amssymb,amsmath,amsthm}
\usepackage[english]{babel}
\usepackage[utf8]{inputenc}

\usepackage{xspace}
\usepackage{tikz}
\usetikzlibrary{arrows,decorations,positioning,shapes}
\usepackage{setspace}
\usepackage{enumitem}
\setlist{nolistsep}
\usepackage{listings}

\usepackage{textcomp}
\usepackage{stmaryrd}
\usepackage{pifont}
\usepackage{soul}

\usepackage{booktabs}   % for nicer-looking tables
\usepackage{multirow}   % for multirow tables
\usepackage{cite}       % for carrying citations to the next line
\usepackage{subcaption} % for dealing with subfigures
\usepackage{todonotes}  % for nicer-looking todo notes
\usepackage{tablefootnote}
\usepackage[most]{tcolorbox}

\let\llncssubparagraph\subparagraph
\let\subparagraph\paragraph
\usepackage{titlesec}
\let\subparagraph\llncssubparagraph

\usepackage[pdftex%
,colorlinks=true       % false: boxed links; true: colored links
,linkcolor=midblue        % color of internal links
,citecolor=midblue        % color of links to bibliography
,filecolor=black       % color of file links
,urlcolor=midblue         % color of external links
,linktoc=page           % only page is linked
,plainpages=false]{hyperref}
\addto\extrasenglish{%

}
\usepackage{enumitem}
\usepackage{caption} %[font=small,skip=0pt]

\fontfamily{ptm}
\setlist{nolistsep}
\setitemize{noitemsep,topsep=3pt,parsep=2pt,partopsep=2pt}
\setenumerate{noitemsep,topsep=3pt,parsep=2pt,partopsep=2pt}

\titlespacing{\section}{0pt}{*2.5}{*1.0}
\titlespacing{\subsection}{0pt}{*1.25}{*0.75}
\titlespacing{\subsubsection}{0pt}{*1.0}{*0.5}
\makeatletter
\renewcommand\paragraph{\@startsection{paragraph}{4}{\z@}%
                                    {1.0ex \@plus0.35ex \@minus.15ex}%
                                    {-1em}%
                                    {\normalfont\normalsize\bfseries}}

\makeatother

\makeatletter
\def\thm@space@setup{%
  \thm@preskip=1.5pt \thm@postskip=1.5pt
}
\makeatother

\makeatletter
\renewenvironment{proof}[1][\proofname]{\par
  \vspace{-0.25\topsep}% remove the space after the theorem
  \pushQED{\qed}%
  \normalfont
  \topsep0pt \partopsep0pt % no space before
  \trivlist
  \item[\hskip\labelsep
        \itshape
    #1\@addpunct{.}]\ignorespaces
}{%
  \popQED\endtrivlist\@endpefalse
  \addvspace{1pt plus 1pt} % some space after
}
\makeatother

\titleformat{\paragraph}[runin]
            {\normalfont\normalsize\bfseries}{\theparagraph}{0.4em}{}

\theoremstyle{remark}

\theoremstyle{plain}

\newcommand{\tn}{\textnormal}

\newcommand{\vars}{\mathsf{vars}} % variables of formula

\newcommand{\fml}[1]{{\mathcal{#1}}}

\newcommand{\sd}{\mathsf{SD}}
\newcommand{\comps}{\mathsf{Comps}}
\newcommand{\obs}{\mathsf{Obs}}
\newcommand{\diag}{\mathrm{\Delta}}

\newcommand{\pspec}{\langle\sd,\comps,\obs\rangle}
\newcommand{\cnf}{\mathsf{CNF}}

\newcommand{\ab}{\mathsf{Ab}}

\newcommand{\sat}{\textsf{SAT}\xspace}
\newcommand{\mxsat}{\textsf{MaxSAT}\xspace}
\newcommand{\wmxsat}{\textsf{WMaxSAT}\xspace}

\newcommand{\minhs}{\textsf{MinHS}\xspace}
\newcommand{\tohit}{\mathbb{U}\xspace}
\newcommand{\mustblock}{\mathbb{D}\xspace}

\definecolor{gray}{rgb}{.4,.4,.4}
\definecolor{midgrey}{rgb}{0.5,0.5,0.5}
\definecolor{middarkgrey}{rgb}{0.35,0.35,0.35}
\definecolor{darkgrey}{rgb}{0.3,0.3,0.3}
\definecolor{darkred}{rgb}{0.7,0.1,0.1}
\definecolor{midblue}{rgb}{0.2,0.2,0.7}
\definecolor{darkblue}{rgb}{0.1,0.1,0.5}
\definecolor{defseagreen}{cmyk}{0.69,0,0.50,0}

\newcommand{\jnoteF}[1]{}

\newcounter{Comment}[Comment]
\DeclareMathOperator*{\nentails}{\nvDash}
\DeclareMathOperator*{\entails}{\vDash}

\DeclareMathSymbol{\Delta}{\mathalpha}{operators}{1}
\DeclareMathSymbol{\Theta}{\mathalpha}{operators}{2}
\DeclareMathSymbol{\Pi}{\mathalpha}{operators}{5}
\DeclareMathSymbol{\Sigma}{\mathalpha}{operators}{6}

\newboolean{extended}      % True if generating the long version of the paper
\setboolean{extended}{false} %
\newcommand{\iflongpaper}[1]{\ifthenelse{\boolean{extended}}{#1}{}}
\newcommand{\ifregularpaper}[1]{\ifthenelse{\boolean{extended}}{}{#1}}

\newboolean{addthanks}      % True if including acknowledgements
\setboolean{addthanks}{false} %
\newcommand{\ifaddack}[1]{\ifthenelse{\boolean{addthanks}}{#1}{}}

\newcommand{\SAT}{\mathsf{SAT}}

{\noindent \textit{Proof sketch#1.}}{\mbox{}\nobreak\hfill\hspace{6pt}$\Box$}

\lstdefinelanguage{RuleList}
                  {morekeywords={IF,THEN,AND}, sensitive=true,
                    morecomment=[l]{;;} }

\newcommand{\mailtodomain}[1]{\href{mailto:#1@ciencias.ulisboa.pt}{\nolinkurl{#1}}}

\usetikzlibrary{circuits.logic.US}

\usetikzlibrary{calc}

\usetikzlibrary{arrows,decorations.pathmorphing,decorations.footprints,fadings,calc,trees,mindmap,shadows,decorations.text,patterns,positioning,shapes,matrix,fit}
\usetikzlibrary{arrows,shadows,backgrounds}
\usetikzlibrary{positioning,fit}
\usetikzlibrary{automata}
\usetikzlibrary{matrix}
\usetikzlibrary{shapes.symbols,shapes.misc,shapes.arrows}
\usetikzlibrary{matrix,chains,scopes,decorations.pathmorphing}

\graphicspath{{plots/}}
\DeclareGraphicsExtensions{.pdf}

\def\jpms{Joao Marques-Silva}
\def\alexi{Alexey Ignatiev}
\def\ajrm{Antonio Morgado}

\title{Model Based Diagnosis of Multiple Observations \\ with Implicit Hitting Sets}
\titlerunning{MBD with Multiple Observations}

\author{
  {\alexi}\inst{1,2}
  \and
  {\ajrm}\inst{1}
  \and
  {\jpms}\inst{1}
}
\authorrunning{Ignatiev, Morgado \& Marques-Silva}

\institute{%
  LASIGE, Faculty of Science, University of Lisbon, Portugal\\
  {{\{\mailtodomain{aignatiev}\texttt{,}\mailtodomain{ajmorgado}\texttt{,}\mailtodomain{jpms}\}\texttt{@ciencias.ulisboa.pt}}}\\
  \and
  ISDCT SB RAS, Irkutsk, Russia
}

\begin {document}
\maketitle
\setcounter{footnote}{0}

%------------------------------------------------------------------------------%
%------------------------------------------------------------------------------%
% File:        abs.tex
%
% Description: 
%
% Created:     10 Apr 2017.
%
% Author:      Joao Marques-Silva (jpms).
%------------------------------------------------------------------------------%
%
\begin{abstract}
  Model based diagnosis finds a growing range of practical
  applications, and significant performance-wise improvements 
  have been achieved in recent years. Some of these improvements
  result from formulating the problem with maximum satisfiability
  (MaxSAT).
  Whereas recent work focuses on analyzing failing observations
  separately, it is also the case that in practical settings there
  may exist many failing observations.
  This paper first investigates the drawbacks of analyzing failing
  observations separately. It then shows that existing solutions do
  not scale for large systems. %with many failing observations,
  Finally, the paper proposes a novel approach for diagnosing systems
  with many failing observations.
  The proposed approach is based on implicit hitting sets and so is
  tightly related with the original seminal work on model based
  diagnosis.
  The experimental results demonstrate not only the importance of
  analyzing multiple observations simultaneously, but also the
  significance of the implicit hitting set approach.
\end{abstract}
%
%------------------------------------------------------------------------------%

%------------------------------------------------------------------------------%
%
%------------------------------------------------------------------------------%
% File:        intro.tex
%
% Description: 
%
% Created:     10 Apr 2017.
%
% Author:      Joao Marques-Silva (jpms).
%------------------------------------------------------------------------------%

\section{Introduction} \label{sec:intro}

The problem of model-based diagnosis~\cite{reiter-aij87} (MBD) is
ubiquitous in practical settings, ranging from the diagnosis of
mechanical and hardware systems, to software programs, to end-user
software (e.g.\ spreadsheets), to knowledge representation systems
(e.g.\ ontologies, etc.), to logic programs, to production systems, to
databases, to triple stores, among many
others~\cite{majumdar-pldi11,roychoudhury-cacm16,wotawa-tse16,jannach-asej16,chomicki-tlp03,schaub-tlp11}.

The theoretical underpinnings of MBD were developed in the mid
80s~\cite{reiter-aij87,reiter-aij92}, and a large body of significant work
followed, covering different approaches for MBD~\cite{darwiche-jair98,veneris-tcad05,bauer-cpaior05,williams-dam07,siddiqi-ijcai07,veneris-fmcad07,dekleer-dxc09,provan-dx10,codish-aaai12,stern-aaai12,wotawa-dx12,wotawa-ijcai13,codish-jair14,msjim-ijcai15,roychoudhury-cacm16}.
In recent years, research has focused on approaches for computing
minimum-size diagnoses, with \mxsat algorithms (and variants thereof)
shown to outperform other approaches in representative
settings~\cite{codish-aaai12,codish-jair14,msjim-ijcai15}.
Another line of work has been on computing (many) subset-minimal
diagnoses~\cite{wotawa-ijcai13}. However, complete enumeration of all
subset-minimal diagnosis is in general
infeasible~\cite{wotawa-ijcai13}, which explains the interest in
computing minimum-size diagnoses.

The usual formulation of MBD is~\cite{reiter-aij87}: given a system
description, composed of some components, where some of these
components can be faulty, and an observation inconsistent with the
system description, select a cardinality-minimal (or subset-minimal)
set of components which, if declared faulty (i.e.\ any behavior is
allowed for the component), then consistency between the model and the
observation is reached.
This formulation of MBD is well-suited in settings where the goal is
to investigate a single (failing) observation. However, the standard
formulation of MBD can be less adequate in recent practical
instantiations of the problem, where one may need to investigate
\emph{many} failing observations and not only a single one. This is
the case with software fault
localization~\cite{majumdar-pldi11,roychoudhury-cacm16,wotawa-tse16,nakajima-jip16}
and spreadsheet debugging~\cite{jannach-asej16}, among others. For
example, for software fault localization, one may be faced with a few
hundred (or even thousands) observations~\cite{rothermel-ese05}.
To our best knowledge, research on diagnosing multiple (failing)
observations is scarce, and existing solutions are not only
unrealistic in practice, but also technically problematic, as
described in this paper.

This paper proposes a novel approach to diagnosing multiple failing
observations concurrently, in such a way that the observed drawbacks
of alternatively solutions are addressed.
The proposed approach builds on recent work on implicit hitting set
dualization~\cite{karp-cpm10,karp-soda11,bacchus-cp11,stern-aaai12,karp-or13,liffiton-cpaior13,lpmms-cj16,jarvisalo-sat16},
and is shown to not only perform efficiently in practice, but also to
overcome the key issue with problem representation size for large
number of observations.
Nevertheless, MBD represents a formidable task, and the solution we
propose is part of a continued effort for developing effective
solutions to aid in diagnosing practical systems.

The paper is organized as follows.~\autoref{sec:prelim} introduces the
notation and definitions used throughout.~\autoref{sec:motiv}
motivates the importance of MBD for multiple observations. The
following two sections investigate approaches for MBD given multiple
observations.~\autoref{sec:basic} outlines existing and also
straightforward solutions. In contrast,~\autoref{sec:dualiter} details
a novel approach based on implicit hitting set
dualization.~\autoref{sec:res} analyzes preliminary results, intended
to highlight the benefits of using implicit hitting set dualization.%
~\autoref{sec:conc} concludes the paper.

%------------------------------------------------------------------------------%

%------------------------------------------------------------------------------%
% File:        prelim.tex
%
% Description:
%
% Created:     10 Apr 2017.
%
% Author:      Joao Marques-Silva (jpms).
%------------------------------------------------------------------------------%

\section{Preliminaries} \label{sec:prelim}

The paper assumes definitions that are standard in Propositional
Satisfiability (\sat), Maximum Satisfiability
(\mxsat)~\cite{sat-handbook09}, and Model-Based Diagnosis (MBD).
These are reviewed in this section.
%
%These include definitions of clauses, Conjunctive Normal Form (CNF)
%formulas, truth assignments, formula satisfiability, formula
%entailment, minimal unsatisfiability, including minimal unsatisfiable
%subsets (MUSes) and minimum correction subsets (MCSes).
%
%In addition, standard definitions used in \mxsat apply, including the
%fact that MaxSAT solutions denote MCSes of smallest size.
%
%
%The paper builds on recent work in model based diagnosis, and uses
%similar
%notation~\cite{codish-aaai12,stern-aaai12,wotawa-ijcai13,codish-jair14,msjim-ijcai15}.
%
%Moreover, basic knowledge of recent %%core-guided
%\mxsat algorithms, i.e.\ \mxsat algorithms based on the iterative
%identification of unsatisfiable subformulas (or cores), is also
%assumed~\cite{ansotegui-aij13,mhlpms-cj13}.
%
%In addition, standard definitions of circuits are
%assumed~\cite{sat-handbook09}, including controlling
%(resp.\ non-controlling) input values of simple gates, concretely 0
%(resp.\ 1) for AND/NAND and 1 (resp.\ 0) for OR/NOR).

\paragraph{Boolean Satisfiability \& Maximum Satisfiability.}
Propositional variables are taken from a set $X=\{x_1,x_2,\ldots\}$.
A Conjunctive Normal Form (CNF) formula is defined as a conjunction of
disjunctions of literals, where a literal is a variable or its
complement. CNF formulas can also be viewed as sets of sets of
literals, and are represented with calligraphic letters, $\fml{A}$,
$\fml{F}$, $\fml{H}$, etc.
Given a formula $\fml{F}$, the set of variables is
$\vars(\fml{F})\subseteq X$.
A truth assignment $\nu$ is a map from variables to $\{0,1\}$. Given a
truth assignment, a clause is satisfied if at least one of its
literals is assigned value 1; otherwise it is falsified. A formula is
satisfied if all of its clauses are satisfied; otherwise it is
falsified.
If there exists no assignment that satisfies a CNF formula $\fml{F}$,
then $\fml{F}$ is referred to as \emph{unsatisfiable}.
(Boolean) Satisfiability (SAT) is the decision problem for
propositional formulas, i.e.\ to decide whether a given propositional
formula is satisfiable.
%whereas CNFSAT is the decision problem of propositional formulas in
%CNF.
Since the paper only considers propositional formulas in CNF,
throughout the paper SAT refers to the decision problem for
propositional formulas in CNF.
Modern SAT solvers instantiate the Conflict-Driven Clause Learning
paradigm~\cite{sat-handbook09}.
For unsatisfiable (or inconsistent) formulas, MUSes (minimal
unsatisfiable subsets) represent subset-minimal subformulas that are
unsatisfiable (or inconsistent), and MCSes (minimal correction
subsets) represent subset-minimal subformulas such that the complement
is satisfiable~\cite{sat-handbook09}.

The (plain) \mxsat problem is to find a truth assignment that
maximizes the number of satisfied clauses. For the plain \mxsat
problem, all clauses are \emph{soft}, meaning that these may not be
satisfied. Variants of the \mxsat can consider the existence of
\emph{hard} clauses, meaning that these must be satisfied, and also
assign weights to the soft clauses, denoting the \emph{cost} of
falsifying the clause; this is referred as the weighted \mxsat
problem, \wmxsat. When addressing \mxsat problems with weights, hard
clauses are assigned a large weight $\top$.
The notation $(c, w)$ will be used to represent a clause $c$ with $w$
denoting the cost of falsifying $c$. The paper considers partial
MaxSAT instances, with hard clauses, for which $w=\top$, and soft
clauses, for which $w=1$. The notation $\langle\fml{H},\fml{S}\rangle$
is used to denote partial \mxsat problems with sets of hard
($\fml{H}$) and soft ($\fml{S}$) clauses.

\paragraph{Model-Based Diagnosis.}
The paper considers standard model-based diagnosis (MBD)
definitions, following Reiter's seminal work~\cite{reiter-aij87}, and
which are used in most modern
references~\cite{reiter-aij87,siddiqi-ijcai11,codish-aaai12,wotawa-ijcai13,codish-jair14,msjim-ijcai15}.
As in recent MBD work, the weak fault model (WFM) is assumed
throughout.
A system description $\sd$ is a set of first-order
sentences~\cite{reiter-aij87}. The system components, $\comps$, are a
set of constants, $\comps=\{c_1,\ldots,c_m\}$.
Given a system description $\sd$, composed of a set of components
$\comps$, each component can be declared as {\em healthy} or {\em
  unhealthy}.
For each component $c\in\comps$, $\ab(c)=1$ if $c$ is declared as
\emph{unhealthy} (or \emph{abnormal}); otherwise $\ab(c)=0$.
Similarly to earlier
work~\cite{provan-dx10,codish-aaai12,codish-jair14,msjim-ijcai15}, it
is assumed that $\sd$ is represented as a CNF formula, namely:
\begin{equation} \label{eq:sd-def}
  \sd\triangleq\bigwedge_{c\in\comps}\,(\ab(c)\lor\fml{F}_c)
\end{equation}
where $\fml{F}_{c}$ denotes the CNF encoding of component $c$.

Observations are used to represent situations where the behavior of
the system is not the expected one. An observation $\obs$ is defined
as a finite set of first-order sentences~\cite{reiter-aij87}.
As with the system description, it is assumed that the observation can
be encoded into CNF, as a set of unit clauses, and denoted $\obs$.

%%\begin{comment}
%
%
\begin{definition}[Diagnosis Problem] \label{def:dp}
A system with description $\sd$ is faulty if it is inconsistent with a
given observation $\obs$ when all components are declared healthy:
\begin{equation} \label{eq:dp}
\sd\land\obs\land\bigwedge_{c\in\comps}\neg\ab(c)\,\entails\bot
\end{equation}
The problem of diagnosis is to identify a set of components which, if
declared unhealthy, make the system consistent with the observation.
The problem of %%model-based diagnosis (
MBD is represented by the 3-tuple $\pspec$.
\end{definition}

\begin{definition}[Diagnosis] \label{def:diag}
Given an MBD problem $\pspec$, the set of components $\diag\subseteq\comps$
is a {\em diagnosis} if
\begin{equation} \label{eq:diag}
\sd\land\obs\land\bigwedge_{c\in\diag}\ab(c)\land\bigwedge_{c\in\comps\setminus\diag}\neg\ab(c)\,\nentails\bot
\end{equation}
A diagnosis $\diag$ is {\em minimal} if no proper subset
$\diag'\subsetneq\diag$ is a diagnosis, and $\diag$ is of {\em minimal
  cardinality} if there exists no other diagnosis
$\diag'\subseteq\comps$ with $|\diag'|<|\diag|$.
\end{definition}
%
%
%%\end{comment}

In this paper, the dual of a diagnosis will be referred to as an
\emph{explanation}. (These are often referred to as
\emph{conflicts}~\cite{reiter-aij87}.) It is well-known that a minimal
diagnosis is a minimal hitting set of the minimal explanations, and
vice-versa~\cite{reiter-aij87}.

%%\subsection{Model-Based Diagnosis with SAT/\mxsat}
%%\subsection{MBD with SAT \& \mxsat}

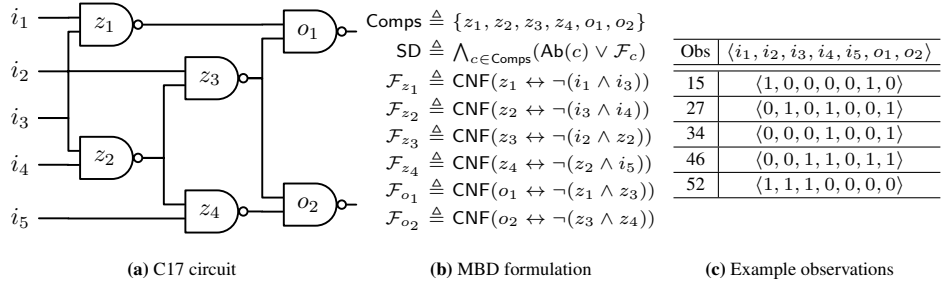
\begin{figure*}[t]
  \begin{subfigure}[b]{0.3625\textwidth}
    \hspace*{-0.425cm}
    \scalebox{0.9125}{
      %------------------------------------------------------------------------------%
% File:        circ2.tex
%
% Description: 
%
% Created:     27 Jun 2014.
%
% Author:      Joao Marques-Silva (jpms).
%------------------------------------------------------------------------------%

\begin{tikzpicture}
  [circuit logic US,line width=0.8pt,line cap=round,line join=round]

  \matrix[column sep=6mm] {
    \node (i1) {$i_1$}; & \node [nand gate,yshift=-1mm] (z1) {$z_1$};
    & &[-1mm] \node [nand gate,yshift=-2mm] (o1) {$o_1$};
     &[-4.5mm] \node[yshift=-2mm] (o1o) {};
    \\
    \node (i2) {$i_2$}; & &[-1mm] \node [nand gate,yshift=-1mm] (z3) {$z_3$};
    &
    &
    \\
    \node [yshift=-2mm] (i3) {$i_3$}; & & &
    &
    \\
    \node [yshift=-2mm] (i4) {$i_4$}; & \node [nand gate,yshift=-1mm] (z2) {$z_2$};
    & &
    &
    \\
    \node [yshift=-1mm] (i5) {$i_5$}; & &[-1mm] \node [nand gate] (z4) {$z_4$};
    &[-1mm] \node [nand gate,yshift=1mm] (o2) {$o_2$};
    &[-4.5mm] \node[yshift=1mm] (o2o) {};
    \\
  };

  \draw
  % z1 gate inputs
  (i1) -- ++(right:7mm) |- (z1.input 1)
  (i3) -- ++(right:7mm) |- (z1.input 2)
  % z2 gate inputs
  (i3) -- ++(right:7mm) |- (z2.input 1)
  (i4) -- ++(right:7mm) |- (z2.input 2)
  % z3 gate inputs
  (i2) -- ++(right:7mm) |- (z3.input 1)
  (z3.input 2) -- ++(left:3mm) |- (z2.output)
  % z4 gate inputs
  (z4.input 1) -- ++(left:3mm) |- (z2.output)
  (i5) -- ++(right:7mm) |- (z4.input 2)
  % o1 gate inputs
  (o1.input 1) -- ++(left:3mm) |- (z1.output)
  (o1.input 2) -- ++(left:3mm) |- (z3.output)
  % o2 gate inputs
  (o2.input 1) -- ++(left:3mm) |- (z3.output)
  (o2.input 2) -- ++(left:3mm) |- (z4.output)
  % Outputs
  (o1.output) -- (o1o) node[near end,above] {}
  (o2.output) -- (o2o) node[near end,above] {}
;
\end{tikzpicture}

%------------------------------------------------------------------------------%
    }
    %%\vspace*{0.3125cm}
    %%\vspace*{0.15cm}
    \caption{C17 circuit}
  \end{subfigure}
  \begin{subfigure}[b]{0.34\textwidth}
    %\vspace*{-0.75cm}
    {\scriptsize
      \[
      \begin{array}{rcl}
        \comps&\triangleq&\{z_1,z_2,z_3,z_4,o_1,o_2\}\\[2pt]
        \sd&\triangleq& \bigwedge_{c\in\comps}(\ab(c)\lor\fml{F}_{c})\\[2pt]
        %\obs&\triangleq&\{i_1,i_2,i_3,\neg i_4,i_5,\neg o_1,\neg o_2\}\\[3pt]
        \fml{F}_{z_1}&\triangleq&\cnf(z_1\leftrightarrow\neg(i_1\land i_3))\\[1.0pt]
        \fml{F}_{z_2}&\triangleq&\cnf(z_2\leftrightarrow\neg(i_3\land i_4))\\[1.0pt]
        \fml{F}_{z_3}&\triangleq&\cnf(z_3\leftrightarrow\neg(i_2\land z_2))\\[1.0pt]
        \fml{F}_{z_4}&\triangleq&\cnf(z_4\leftrightarrow\neg(z_2\land i_5))\\[1.0pt]
        \fml{F}_{o_1}&\triangleq&\cnf(o_1\leftrightarrow\neg(z_1\land z_3))\\[1.0pt]
        \fml{F}_{o_2}&\triangleq&\cnf(o_2\leftrightarrow\neg(z_3\land z_4))\\%%[1.0pt]
      \end{array}
      \]
    }
    %%\vspace*{0.15cm}
    \vspace*{-0.2cm}
    \caption{MBD formulation}
  \end{subfigure}
  \begin{subfigure}[b]{0.275\textwidth}
    %\vspace*{-0.75cm}
    %\vspace*{0.05cm}
    %% \centering
    {\scriptsize%\footnotesize
      
      \begin{tabular}{c|c} %{|c|c|}
        \hline
        Obs & $\langle i_1,i_2,i_3,i_4,i_5,o_1,o_2\rangle$ %& $\langle\rangle$
        \\ \hline\hline
        15 & $\langle1,0,0,0,0,1,0\rangle$ %& $\langle\rangle$
        \\ \hline
        27 & $\langle0,1,0,1,0,0,1\rangle$ %& $\langle\rangle$
        \\ \hline
        34 & $\langle0,0,0,1,0,0,1\rangle$ %& $\langle\rangle$
        \\ \hline
        46 & $\langle0,0,1,1,0,1,1\rangle$ %& $\langle\rangle$
        \\ \hline
        52 & $\langle1,1,1,0,0,0,0\rangle$ %& $\langle\rangle$
        \\ \hline
        %
        %61 & $\langle0,0,1,0,1,1,0\rangle$ & $\langle\rangle$
        %\\ \hline
        %
      \end{tabular}
    }
    \vspace*{0.575cm}
    \caption{Example observations}
  \end{subfigure}
  \caption{C17 circuit and selected observations from ISCAS85 scenarios}
  \label{fig:c17-ex}
\end{figure*}

To model MBD with \mxsat~\cite{veneris-fmcad07,provan-dx10}, $\sd$
(see \eqref{eq:sd-def}) represents the {\em hard} clauses, whereas the
{\em soft clauses} are unit clauses $(\neg \ab(c))$, one for each
component $c\in\comps$.
%
%This is referred to as the {\em basic} \mxsat encoding in this paper.
Different \mxsat solving approaches can then be applied.
Alternatively, the soft clauses can be replaced by a cardinality
constraint and solved iteratively with a SAT solver.
Combinational circuits represent the most often used vocabulary in
MBD-related research~\cite{reiter-aij87}.
\autoref{fig:c17-ex} illustrates an example circuit, example
observations, and an often used encoding into
CNF~\cite{veneris-tcad05,provan-dx10,wotawa-ijcai13,msjim-ijcai15}.
An alternative model, requiring more clauses, has also been studied in
recent times~\cite{codish-aaai12,codish-jair14}. In this paper we
follow the original simpler model.

Moreover, although combinational circuits have often been used as the
main vehicle to convey research ideas in MBD, other vocabularies can
be used. One example is of course clauses, i.e.\ each component is a
clause. In this paper, the key ideas will be conveyed by system
descriptions where each component is a clause. Additional examples,
using different vocabularies, are also analyzed throughout the paper.

% put back in the final version
% The paper explores implicit (minimal) hitting sets, as applied
% recently in different
% settings~\cite{karp-cpm10,karp-soda11,bacchus-cp11,stern-aaai12,karp-or13,liffiton-cpaior13,pms-aaai13,lpmms-cj16}.
%
The paper explores implicit (minimal) hitting sets, as applied
recently in different
settings~\cite{karp-cpm10,karp-soda11,bacchus-cp11,stern-aaai12,karp-or13,liffiton-cpaior13,lpmms-cj16}.
For the concrete case of MBD, the paper also relates implicit minimal
hitting sets and the duality between minimal diagnoses and minimal
explanations~\cite{reiter-aij87,stuckey-padl05}.

\paragraph{Relating MBD with MCSes \& MUSes.}
It is important to highlight that there is a close relationship
between diagnoses and MCSes, and between explanations and
MUSes~\cite{reiter-aij87,lozinskii-jetai03,stuckey-padl05,msjim-ijcai15}.
Indeed, given the inconsistent formula~\eqref{eq:sd-def}, a minimal
diagnosis $\diag$ is such that~\eqref{eq:diag} is consistent. Thus,
$\diag$ is an MCS of~\eqref{eq:sd-def}.
Similarly, an explanation is a minimal hitting set of the diagnoses,
and so it corresponds to an MUS of~\eqref{eq:sd-def}.
As a result, enumeration of diagnoses can be obtained by enumeration
of MCSes~\cite{mshjpb-ijcai13}, and enumeration of explanations by
enumeration of MUSes~\cite{lpmms-cj16}.
Given the above, and throughout this paper, the term MCS is used
interchangeably with minimal diagnosis, and the term MUS is used
interchangeably with minimal explanation.

\paragraph{Multiple Observations.}
The MBD problem can be generalized to the situation where multiple
inconsistent observations exist.
In the presence of multiple observations,~\eqref{eq:diag} is modified
as follows for observation $i$, $\obs_i$:
\begin{equation} \label{eq:diagmo}
\sd_i\land\obs_i\land\bigwedge_{c\in\diag}\ab(c)\land\bigwedge_{c\in\comps\setminus\diag}\neg\ab(c)\,\nentails\bot
\end{equation}
We assume that the system remains unchanged given different
observations, and so $\sd_i$ is solely a \emph{replica} of the system
description $\sd$. A more general setting, in which the system
considered also changes with the observation, could be considered, but
would not change the main results in the paper.
Observe that we need a distinct replica for each observation, since the
actual values that result given the observation may differ.

\begin{comment}
Given $r$ observations, we can consider the following aggregated
problem formulation:
%
{\small
\begin{equation} \label{eq:diagmo2}
\bigwedge_{i=1}^{r}\left(\sd_i\land\obs_i\land\bigwedge_{c\in\diag}\ab(c)\land\bigwedge_{c\in\comps\setminus\diag}\neg\ab(c)\right)\,\nentails\bot
\end{equation}
}
%
where the goal is to find a subset-minimal (or cardinality-minimal)
diagnosis $\Delta$ that makes the system consistent for \emph{any}
observation.
\end{comment}

%\begin{figure*}[t]
%  \centering \includegraphics[scale=0.75]{./figs/exampl02.pdf}
%  \caption{System for running example} \label{fig:exampl2}
%\end{figure*}

\begin{definition}[MBD with Multiple Observations]
  We assume a sequence of observations $\obs_i$, with $1\le i\le r$.
  With each observation we associate a replica of the system $\sd_i$,
  but such that the abnormal variables are shared by the different replicas.
  A minimal diagnosis $\Delta\subseteq\comps$ is a minimal set such
  that,
  {\small
    \begin{equation} \label{eq:diagmo2}
      \bigwedge_{i=1}^{r}\left(\sd_i\land\obs_i\right)\land\bigwedge_{c\in\diag}\ab(c)\land\bigwedge_{c\in\comps\setminus\diag}\neg\ab(c)\,\nentails\bot
    \end{equation}
  }
  holds.
\end{definition}
Thus, the goal is to find a subset-minimal (or possibly a
cardinality-minimal) diagnosis $\Delta\subseteq\comps$ that makes the
system consistent with \emph{any} of the observations $\obs_i$,
$1\le i\le r$.

%------------------------------------------------------------------------------%

%------------------------------------------------------------------------------%
% File:        motiv.tex
%
% Description:
%
% Created:     10 Apr 2017.
%
% Author:      Joao Marques-Silva (jpms).
%------------------------------------------------------------------------------%

\section{The Need for Multiple Observations} \label{sec:motiv}

The formalization of model-based diagnosis presented in the previous
section reveals fundamental challenges for MBD in practical settings.
Concretely, MBD can be viewed as the process of identifying plausible
(usually subset- or cardinality-minimal) guesses of which components
in a system must be declared abnormal for consistency to be attained,
when the abnormal components are allowed \emph{any}
behaviors. However, such guesses may not represent actual faulty
components.
The goal of MBD is to achieve consistency, but this is often possible
by declaring faulty components that are unrelated with the actual
bug.
More importantly, complete enumeration of all diagnoses is infeasible
in practice~\cite{wotawa-ijcai13}. As a result, computed diagnoses
should be as accurate as possible, given the information about failing
observations.

This section investigates the importance of considering multiple
observations to increase the accuracy of computed diagnoses.
As shown below, it is simple to find failing observations which
mislead the diagnosis tool, such that an exponentially large number of
inaccurate and so irrelevant diagnoses are generated before the
correct diagnosis is computed. More, importantly, this situation
occurs in essentially endless situations.
By considering multiple observations the accuracy of MBD can only
improve. Whereas it is in general difficult to identify a test which
will reveal as the source of the bug only the bug location, one can
expect multiple tests to elicit cooperation such that only actual
fault locations are reported.

\paragraph{A Buggy CNF Encoder.}
A well known example for SAT researchers is the debugging of CNF
encoders. Suppose one implements an encoder to represent in CNF some
problem. Suppose further that this encoder is composed on some modules
(say $M_1$, $M_2$, $M_3$ and $M_4$), and that some have been tested
and used before (e.g.\ $M_1$ and $M_3$) and that some others have
recently been implemented (e.g.\ $M_2$ and $M_4$).
Let us consider the example formula below.
\begin{equation} \label{eq:cnf:ex01}
  %%\small
  \begin{array}{rl}
    \{ &
    (\neg x_{11}\lor y_{2a},\top), (\neg x_{12}\lor y_{2a},\top),\ldots,
    (\neg x_{1r-1}\lor y_{2a},\top), \\
    & (\neg y_{2a}\lor\neg t_{21a}\lor y_{21b},1), (\neg y_{21b}\lor\neg t_{21a}\lor y_{21c},1),\\
    & (\neg y_{21c}\lor\neg t_{21a}\lor y_{21d},1), (\neg y_{21d}\lor\neg t_{21a}\lor w_{42a},1), \\
    & \ldots \\
    & (\neg y_{2a}\lor\neg t_{21a}\lor y_{2kb},1), (\neg y_{2kb}\lor\neg t_{21a}\lor y_{2kc},1),\\
    & (\neg y_{2kc}\lor\neg t_{21a}\lor y_{2kd},1), (\neg y_{2kd}\lor\neg t_{21a}\lor w_{41a},1), \\
    & (\neg s_{31a}\lor w_{41a},\top), (\neg s_{31a}\lor w_{42a},\top), \\
    & (\neg w_{41a}\lor u_{41a},\top),
    (\neg w_{42a}\lor u_{42a},\top), \\
    & (\neg u_{41a}\lor\neg t_{41a}\lor z_{41a},1), (\neg u_{42a}\lor\neg t_{41a}\lor z_{42a},1) \} \\
  \end{array}
\end{equation}
Some clauses are marked as hard since these result from modules of the
encoder we know are not buggy, e.g.\ $M_1$ and $M_3$.
Some other clauses are generated by modules which we do not know
whether they are buggy, e.g.\ $M_2$ and $M_4$. For the example shown,
this is the case with the clauses containing variables from the sets
of $\{y, t\}$ variables, or from the sets of $\{y, t, w\}$ variables,
or from the sets of $\{t, u, z\}$ variables.
Concretely, for $M_4$, the clauses only have literals in the set of
$\{t, u, z\}$ variables (i.e.\ these are the last two clauses above).
All the other soft clauses are produced by $M_2$.
Let us assume further that the expected behavior of the CNF encoder is
summarized in~\eqref{eq:tst:ex01} below. Each line can be viewed as an
observation (or as a \emph{failing test}), since the assignments
reported in each line are \emph{inconsistent} with the model of the
system given by~\eqref{eq:cnf:ex01}.
\begin{equation} \label{eq:tst:ex01}
  %%\begingroup
  %%\renewcommand*{\arraystretch}{1.5}
  \setlength\arraycolsep{5pt}
  \begin{vmatrix}
    \tn{test} & x_{11} &x_{12}&\cdots &x_{1r-1}&s_{31a}&t_{21a}&t_{41a}&z_{41a}&z_{42a} \\
    t_1      & 1     & 0  &\cdots & 0    & 0    & 1     & 1    & 0    & 0    \\
    t_3      & 0     & 1  &\cdots & 0    & 0    & 1     & 1    & 0    & 0    \\
    \vdots   &\vdots&\vdots&\vdots& 0    & 0    & 1     & 1    & 0    & 0    \\
    t_{r-1} & 0     & 0    &\cdots & 1    & 0    & 1     & 1    & 0    & 0    \\
    t_{r} & 0  & 0    &\cdots & 0  & 1    & 1     & 1    & 0    & 0    \\
  \end{vmatrix}
  %%\endgroup
\end{equation}
More importantly, a quick inspection suggests that by declaring the
clauses with the $w$, $u$  and $z$ variables as faulty, i.e.\ the last
two clauses, we can make the rest of the formula consistent with
\emph{all} the tests.
Unfortunately, if we decide to analyze each test separately, starting
with $t_1$, then each of the first $r-1$ tests will produce
$4^{k} + 4^{\lfloor k/2\rfloor} + 4^{\lceil k/2\rceil} + 1$
diagnoses, of which
$4^{k} + 4^{\lfloor k/2\rfloor} + 4^{\lceil k/2\rceil}$
do \emph{not} represent a valid diagnosis for all the tests, given
$t_{r}$.

Clearly, there exists a \emph{single} MCS, involving 2 clauses which,
if removed, causes~\eqref{eq:cnf:ex01} to become consistent with
\emph{all} tests. This is the MCS we are interested in. However, if we
opt to analyze each test separately, and for large enough $k$, it will
be unrealistic to compute the correct MCS.

%Once more, by devising a number of tests and running a model based
%diagnosis tool, we will compute an exponentially large number of
%diagnoses, even though it is plain that the problem is in the last two
%clauses.
%%
%For this concrete example, let us assume that the test cases are the
%following ones:
%%
%For each of the first $r$ test cases, the clauses involving the $y$
%variables will yield $4^k+1$ MCSes.
%%
%In constrast, the last test will one MCS.
%%
%More importantly, the MCS obtained with the last test represents a
%diagnosis that correctly identifies the source of the bug.

\paragraph{Detailed Analysis.}
For completeness, we derive the number of diagnoses indicated above
for the ``buggy'' formula~\eqref{eq:cnf:ex01}, given the set of
tests~\eqref{eq:tst:ex01}.

To make the system consistent we can declare faulty the last two
clauses:
\begin{equation}
  \begin{array}{rcl}
    \{ & (\neg u_{41a}\lor\neg t_{41a}\lor z_{41a},1), (\neg
    u_{42a}\lor\neg t_{41a}\lor z_{42a},1) & \}
  \end{array}
\end{equation}
This represents the diagnosis common to all tests.
However, for the first $r-1$ tests, we also need to consider the
groups of four clauses:
\begin{equation}
  \begin{array}{rll}
    \{ & (\neg y_{2a}\lor\neg t_{21a}\lor y_{2jb},1),
         (\neg y_{2jb}\lor\neg t_{21a}\lor y_{2pc},1), \\
    & (\neg y_{2pc}\lor\neg t_{21a}\lor y_{2pd},1),
    (\neg y_{2pd}\lor\neg t_{21a}\lor w_{4qa},1) & \} \\
  \end{array}
\end{equation}
with $1\le p\le k$ and $1\le q\le 2$, such that $q=1$ for $p$ odd, and
$q=2$ for $p$ even.
Thus, for the first $r-1$ tests there is a diagnosis between each
group of four clauses (i.e.\ each value of $p$) for the two values of
$q$. Moreover, a diagnosis can be obtained by picking one group of
four clauses (for one value of $q$) and one of the final clauses (for
the other value of $q$).
To get the total number of diagnoses for each of the first $r-1$
tests, we just need to aggregate the different contributions.
For the first set of groups of 4 clauses, the contribution is
$4^{k}$ diagnoses.
Between each group of four clauses and the corresponding final single
clause, the contribution is $4^{\lfloor k/2\rfloor} + 4^{\lceil
  k/2\rceil}$ diagnoses.
Finally, the last pair of clauses corresponds to a single diagnosis.

\begin{figure}[!th]
  \begin{subfigure}[t]{0.485\textwidth}
    \scalebox{0.995}{
\begin{lstlisting}[language=RuleList,mathescape,frame=single,basicstyle=\sffamily\footnotesize,lineskip=-1pt]
R11: IF atAndaman
     THEN onHolidays.
R12: IF atBahamas
     THEN onHolidays.
; ...
R1$r-1$: IF atSeychelles
     THEN onHolidays.
R21a: IF onHolidays
      AND luxuryResort
      THEN schedKiteSurf.
R21b: IF schedKiteSurf
      AND luxuryResort
      THEN rentKiteSurf.
R21c: IF rentKiteSurf
      AND luxuryResort
      THEN practiceKiteSurf.
R21d: IF practiceKiteSurf
      AND luxuryResort
      THEN getSomeSleep.
; ...
R2$k$a: IF onHolidays
      AND luxuryResort
      THEN schedWindSurf.
R2$k$b: IF schedWindSurf
      AND luxuryResort
      THEN rentWindSurf.
R2$k$c: IF rentWindSurf
      AND luxuryResort
      THEN practiceWindSurf.
R2$k$d: IF practiceWindSurf
      AND luxuryResort
      THEN getSomeRest.
R31a: IF lateNight
      THEN getSomeRest.
R32a: IF lateNight
      THEN getSomeSleep.
R41a: IF getSomeRest
      AND moreTime
      THEN seeDoctor.
R42a: IF getSomeSleep
      AND moreTime
      THEN callHolidaysOff.
\end{lstlisting}
    }
    \caption{Production system} \label{fig:ex01-rules}
  \end{subfigure}
  \hfill
  \begin{subfigure}[t]{0.485\textwidth}
    \scalebox{0.995}{
\begin{lstlisting}[language=Prolog,frame=single,basicstyle=\sffamily\footnotesize,lineskip=-1pt]
holds(onHolidays) :-
    holds(atAndaman).
holds(onHolidays) :-
    holds(atBahamas).
% ...
holds(onHolidays) :-
    holds(atSeychelles).
holds(schedKiteSurf) :-
    holds(onHolidays),
    holds(luxuryResort).
holds(rentKiteSurf) :-
    holds(schedKiteSurf),
    holds(luxuryResort).
holds(practiceKiteSurf) :-
    holds(rentKiteSurf),
    holds(luxuryResort).
holds(getSomeSleep) :-
    holds(schedKiteSurf),
    holds(luxuryResort).
% ...
holds(schedWindSurf) :-
    holds(onHolidays),
    holds(luxuryResort).
holds(rentWindSurf) :-
    holds(schedWindSurf),
    holds(luxuryResort).
holds(practiceWindSurf) :-
    holds(rentWindSurf),
    holds(luxuryResort).
holds(getSomeRest) :-
    holds(schedWindSurf),
    holds(luxuryResort).
holds(getSomeRest) :-
    holds(lateNight).
holds(getSomeSleep) :-
    holds(lateNight).
holds(seeDoctor) :-
    holds(getSomeRest),
    holds(moreTime).
holds(callHolidaysOff) :-
    holds(getSomeSleep),
    holds(moreTime).

\end{lstlisting}
    }
    \caption{Datalog program} \label{fig:ex01-datalog}
  \end{subfigure}
  \caption{Examples of a production system and a Datalog program}
  \label{fig:ex01}
\end{figure}

\paragraph{An Example with Lists of Rules.}
\autoref{fig:ex01-rules} shows the rules of a production
system\footnote{A simplified propositional version of a production
  system is considered, to illustrate the key points.} intended to
serve as an assistant in the holidays of some user. The rules are
organized by layers, to simplify the analysis, and these layers are
reflected in the numbers used for the rules. Suppose some user
is spending her holidays at a {\sffamily Luxury Resort} in the idyllic
place of {\sffamily Andaman}, and with more time available to engage
on activities at the resort. As can be concluded, application of the
rules in the production system would conclude that the user must see
the Doctor and call the holidays off.
Manual inspection of the rules in the production system reveals that
the problem lies in rules R41a and R42a which derive an unlikely
conclusion given the premises.
Unfortunately, if the production system is much larger, the manual
analysis of the rules will be unrealistic, and so automatic analysis
needs to be considered.
A number of test cases can be envisioned, a propositional formula can
be derived, and MCSes of this formula can be computed. As highlighted
earlier for the CNF encoder example, it may happen that an
exponentially large number of MCSes is computed, most of which not
reflecting actual MCSes given the complete set of failing
observations.
Perhaps not surprisingly, assuming the right set of productions are
known not to be buggy, we can map this example
into~\eqref{eq:cnf:ex01}, and so the same analysis applies.

\paragraph{An Example with Datalog.}
\autoref{fig:ex01-datalog} proposes a different instantiation of the
same problem, where Datalog is used instead of a production system.
If the facts {\sffamily atAndaman}, {\sffamily LuxuryResort} and
{\sffamily moreTime} are added to the Datalog program, we will infer
the same facts as above, i.e.\ we must see the doctor and must call
the holidays off. Once more, inspection will reveal the problem in
some specific rules, whereas model based diagnosis may yield an
exponentially large number of possible diagnoses given a suitable
set of failing tests.
Once again, assuming the right set of productions are known not to be
buggy, we can map this example into~\eqref{eq:cnf:ex01}, and so the
same analysis applies.

\section{Diagnosis of Multiple Observations} \label{sec:basic}

This section investigates approaches for computing diagnoses in the
presence of multiple failing observations, and considers as working
examples those studied in~\autoref{sec:motiv}, either the buggy CNF
encoder, the production system or the Datalog program (among the many
other similar settings that could be considered). For the purposes of
this section, we consider the (``buggy'') CNF formula
in~\eqref{eq:cnf:ex01}.

\paragraph{Separate analysis and posterior assemblage.}
As shown in \autoref{alg:separate-enum}, one solution for simultaneous
diagnosis of multiple observations is to
generate \emph{all} the diagnoses for each observation and then
compute the diagnoses that correct all observations. This approach has
been investigated by a number of researchers in the recent
past~\cite{nakajima-jip16,majumdar-pldi11}.
\begin{algorithm}[t]
%---------------------------------------------------------%
% Author:      Antonio Morgado [ajrmorgado@gmail.com]     %
% Created:     17 February                                %
%---------------------------------------------------------%

\SetStartEndCondition{ }{}{}%
\SetKwProg{Fn}{def}{\string:}{}
\SetKwFunction{Range}{range}%%
\SetKw{KwTo}{in}\SetKwFor{For}{for}{\string:}{}%
\SetKwIF{If}{ElseIf}{Else}{if}{:}{elif}{else:}{}%
\SetKwFor{While}{while}{:}{}%
\SetKwFor{ForEach}{foreach}{:}{}%
\AlgoDontDisplayBlockMarkers\SetAlgoNoEnd\SetAlgoNoLine%
\LinesNumbered
\DontPrintSemicolon%
\SetKw{KwNot}{not\xspace}
\SetKw{KwAnd}{and\xspace}
\SetKw{KwOr}{or\xspace}
\SetKw{KwBreak}{break\xspace}
\SetKwData{false}{{\small false}}
\SetKwData{true}{{\small true}}
\SetKwData{st}{{\slshape st}}
\SetKwFunction{CNF}{CNF}
\SetKwFunction{SAT}{SAT}
\SetKwInOut{Input}{input}\SetKwInOut{Output}{output}
\SetKwComment{tcpy}{\# }{}
%\SetKwFunction{minhs}{MinHS}
\SetKwFunction{smt}{SMT}
\SetKwFunction{reduce}{Reduce}
\SetKwFunction{encode}{Encode}
\SetKwFunction{getlabels}{MapToLabels}
\SetKwFunction{reportdiags}{ReportDiagnoses}
\SetKwFunction{reportexpl}{ReportExpl}
\SetKwFunction{alldiag}{AllDiagnoses}
\SetKwFunction{diagcomb}{DiagCombine}
\Input{$\sd_1,\ldots,\sd_r,\obs_1,\ldots,\obs_r$}
\Output{$\mathbb{D} = \{\Delta_1,\Delta_2\ldots\}$}
\BlankLine

$(\Gamma_1,\ldots,\Gamma_r) \gets (\emptyset,\ldots, \emptyset)$ \;
\ForEach {$i\in\{1,\ldots,r\}$} {
    $\Gamma_i \gets \alldiag(\sd_i,\obs_i)$ \;
}
$\mathbb{D} \gets \diagcomb(\Gamma_1,\ldots,\Gamma_r)$ \;
$\reportdiags(\mathbb{D})$ \;
\Return
\BlankLine

\caption{Enumeration of minimal diagnoses by separate analysis and
posterior assemblage} \label{alg:separate-enum}
\end{algorithm}

A major draback of this approach is that in some settings, the
enumeration of all the diagnoses for a given observation may be
unrealistic~\cite{wotawa-ijcai13}.
Indeed, as analyzyed in~\autoref{sec:motiv}, for some observations
there may exist an exponentially large number of diagnoses, most of
which will then be discarded.
%
%
%The example in~\autoref{fig:exampl2} illustrates this drawback. If the
%algorithm starts with any of the first $r-1$ observations, each will
%require $2+3^k$ diagnoses, even if the final result will be a single
%diagnosis.
%
Observe that even when diagnoses are computed by decreasing size, it
will still be possible to force this algorithm to only find the
correct diagnosis after computing an exponentially large number of
(useless) diagnoses.
%
%For the example in~\autoref{fig:exampl2}, with $m>3$, the intended
%diagnosis would only be identified after the first $1+3^{k}$ smaller
%diagnoses were computed.
%
In a similar vein, tentative approximations would not necessarily be
effective. For example, an approach based on computing the union of
one smallest size diagnosis (i.e.\ the \mxsat solution) for each
different observation would not necessarily solve the problem, since
the smallest size diagnosis might not be accurate as well.
%Once again, for the example in~\autoref{fig:exampl2}, for $m\ge2$,
%the smallest diagnosis corresponds to the buffer gates at the inputs,
%and each one is different for each observation. It should be noted
%that, for this concrete setting, the computed solution would not even
%include what is the correct diagnosis.

\paragraph{Aggregated analysis.}
Another solution consists of simply generating a model that represents
$r$ copies of the system, one for each observation, and then computing
cardina\-li\-ty-minimal or subset-minimal diagnoses. This corresponds to
computing MCSes or \mxsat solutions of~\eqref{eq:diagmo2}.
The model to be generated essentially encodes~\eqref{eq:diagmo2} as a
\mxsat problem, and either uses an MCS extractor to compute a
subset-minimal diagnoses or a \mxsat solver for computing a
cardinality-minimal diagnosis.

Although the aggregated problem formulation will only compute the
actual subset-minimal (or cardinality-minimal) diagnoses for the set
of observations, as we show in~\autoref{sec:res}, it will be
impractical for all but the smaller examples (or with a small number
of observations), given the number of replicas of the system that need
to be considered.

It should be noted that the examples in~\autoref{sec:motiv} aim at
being as simple as possible. Let such an example be denoted by
$\fml{B}=\langle\fml{H},\fml{S}\rangle$.
In general, $\fml{B}$ will be part of a much larger system (e.g.\ in
fault localization, or spreadsheed debugging, among other examples).
Concretely, the general setting will be
$\fml{B}'=\langle\fml{H}\cup\fml{G}_H,\fml{S}\cup\fml{G}_S\rangle$,
with $\fml{G}=\fml{G}_H\cup\fml{G}_S$ denoting the additional clauses
used for encoding the system, but which are not essential for
highlighting the problems with the accuracy of model-based diagnosis
and the diagnosis of multiple failing observations.
Observe that $\fml{G}$ can be arbitraly large, even when $\fml{B}$ is
as small as the examples described in~\autoref{sec:motiv}.
Thus, for aggregated analysis, $\fml{G}$ will be replicated as many
times as the number of failing observations, and the complete formula
is what an MCS enumerator or a \mxsat solver needs to be able to
analyze.

%As shown later in the paper, this approach will not scale in practice
%if the system has a large representation and/or if the number of
%observations is large.

%%\jnote{Formalize this approach.}

%% \begin{table}[t]
%%   \begin{center}
%%     {\small
%%       \renewcommand{\arraystretch}{1.25}
%%       \renewcommand{\tabcolsep}{0.35em}
%%       \begin{tabular}{c|c} \hline
%%         Obs & $\langle x_1,\ldots,x_{r-1},y,w,s_{31},z_1,\ldots,s_{3m},z_m\rangle$ \\ \hline\hline
%%         $1$ & $\langle 1,0,\ldots,0,1,1,0,0,\ldots,0,0\rangle$ \\ \hline
%%         $2$ & $\langle 0,1,\ldots,0,1,1,0,0,\ldots,0,0\rangle$ \\ \hline
%%         $\cdots$ & $\cdots$ \\ \hline
%%         $r-1$ & $\langle 0,0,\ldots,1,1,1,0,0,\ldots,0,0\rangle$ \\ \hline
%%         $r$ & $\langle 0,0,\ldots,0,0,0,1,0,\ldots,1,0\rangle$ \\ \hline
%%       \end{tabular}
%%       \renewcommand{\arraystretch}{1.0}
%%       \renewcommand{\tabcolsep}{0.5em}
%%     }
%%   \end{center}
%%   \caption{Observations for system in~\autoref{fig:exampl2}} \label{tab:obs}
%% \end{table}

%------------------------------------------------------------------------------%

%------------------------------------------------------------------------------%
% File:        dualiter.tex
%
% Description:
%
% Created:     10 Apr 2017.
%
% Author:      Joao Marques-Silva (jpms).
%------------------------------------------------------------------------------%

\section{Iterative Hitting Set Dualization} \label{sec:dualiter}

This section proposes an alternative approach for computing diagnoses
given a (possibly large) set of observations.
In contrast with the approaches described in the previous section,
and so in contrast with earlier work, the proposed approach is shown
to scale in practice.

% put back in the final version!
The proposed approach hinges on recent work on hitting set
dualization, which has been investigated in different
contexts in recent
years~\cite{karp-cpm10,karp-soda11,bacchus-cp11,stern-aaai12,karp-or13,liffiton-cpaior13,pms-aaai13,iplms-cp15,pimms-ijcai15,lpmms-cj16,ipms-cp16,imms-ecai16,jarvisalo-sat16}. (However,
these ideas can be traced to the seminal work of
Reiter~\cite{reiter-aij87}, and have been studied in different
settings over the years~\cite{stuckey-padl05,liffiton-jar08}, among
others.)

% The solution proposed hinges on approaches for implicit hitting set
% dualization, which have been investigated in different contexts in
% recent
% years~\cite{karp-cpm10,karp-soda11,bacchus-cp11,stern-aaai12,karp-or13,liffiton-cpaior13,lpmms-cj16,jarvisalo-sat16}. (Nevertheless,
% these ideas can also be traced to the seminal work of
% Reiter~\cite{reiter-aij87}, and have been studied in different
% settings over the years, e.g.~\cite{stuckey-padl05,liffiton-jar08}
% among others.)

\begin{algorithm}[t]
  %------------------------------------------------------------------------------%
% File:        alg.tex
%
% Description:
%
% Created:     12 Sep 2014.
%
% Author:      Joao Marques-Silva (jpms).
%------------------------------------------------------------------------------%

\SetStartEndCondition{ }{}{}%
\SetKwProg{Fn}{def}{\string:}{}
\SetKwFunction{Range}{range}%%
\SetKw{KwTo}{in}\SetKwFor{For}{for}{\string:}{}%
\SetKwIF{If}{ElseIf}{Else}{if}{:}{elif}{else:}{}%
\SetKwFor{While}{while}{:}{}%
\SetKwFor{ForEach}{foreach}{:}{}%
\AlgoDontDisplayBlockMarkers\SetAlgoNoEnd\SetAlgoNoLine%
\LinesNumbered
\DontPrintSemicolon%
\SetKw{KwNot}{not\xspace}
\SetKw{KwAnd}{and\xspace}
\SetKw{KwOr}{or\xspace}
\SetKw{KwBreak}{break\xspace}
\SetKwData{false}{{\small false}}
\SetKwData{true}{{\small true}}
\SetKwData{st}{{\slshape st}}
\SetKwFunction{CNF}{CNF}
\SetKwFunction{SAT}{SAT}
\SetKwInOut{Input}{input}\SetKwInOut{Output}{output}

\SetKwComment{tcpy}{\# }{}
\SetCommentSty{commfont}
%\SetKwFunction{minhs}{MinHS}
\SetKwFunction{smt}{SMT}
\SetKwFunction{reduce}{Reduce}
\SetKwFunction{encode}{Encode}
\SetKwFunction{getlabels}{MapToLabels}
\SetKwFunction{reportdiag}{ReportDiag}
\SetKwFunction{reportexpl}{ReportExpl}
\Input{$\sd,\obs_1,\ldots,\obs_r$}
\Output{$\mathbb{D} = \{\Delta_1,\Delta_2\ldots\}$, $\mathbb{U} = \{\fml{U}_1,\fml{U}_2\ldots\}$}
\BlankLine

%\BlankLine
$(\fml{H}_{1},\ldots,\fml{H}_{r},\fml{S})\gets\encode(\sd,\obs_1,\ldots,\obs_r)$
\; %, \quad$i=1,\ldots,r$
$(\mustblock,\tohit) \gets (\emptyset,\emptyset)$\;
%\BlankLine
\While {\true} {
  $(\st,\Delta) \gets \minhs(\tohit,\mustblock)$\tcpy*[r]{find a min
    HS of $\tohit$ s.t.\ $\mustblock$}
  \If {\KwNot$\st$} {
    \KwBreak\;
  }
  \ForEach {$i\in\{1,\ldots,r\}$} {
    $(\st,\kappa) \gets \SAT(\fml{H}_{i}\cup(\fml{S}\setminus\Delta))$\;
    \If {\KwNot$\st$} {
      $\fml{U} \gets \reduce(\kappa)$\label{alg:expl}\tcpy*[r]{
        $\fml{U}$ is MUS of $\fml{H}_{i}\cup(\fml{S}\setminus\Delta)$}
      $\tohit \gets \tohit\cup\{\fml{U}\}$\;
      $\reportexpl(\fml{U})$\tcpy*[r]{report min
        explanation} % $\omega_{\pi_i}(\fml{P})$
      \KwBreak\;
    }
  }
  \Else(\hfill\emph{\small\# if the loop was not broken}) {
    $\mustblock \gets \mustblock\cup\{\Delta\}$
    \tcpy*[r]{block diagnosis $\Delta$}
    $\reportdiag(\Delta)$\tcpy*[r]{report min
      diagnosis} %$\delta_{\pi_1,\ldots,\pi_m}(\fml{P})$
  }
  \ForEach {\label{alg:loop-beg}$i\in\{1,\ldots,r\}$} {
    \If(\hfill\emph{\small\# no more diagnoses exist}) {\KwNot$\SAT(\fml{H}_{i}\cup\mustblock)$} {
      \Return\label{alg:loop-end}\;
    }
  }

}
\Return
\BlankLine
%
%------------------------------------------------------------------------------%

  \caption{Enumeration of minimal diagnoses} \label{alg:diag-enum}
\end{algorithm}

The proposed approach is summarized in~\autoref{alg:diag-enum}.
Each $\Delta_i$ denotes a computed minimal diagnosis, and each
$\fml{U}_j$ denotes a computed minimal explanation.
Although the paper focuses mainly on subset-minimal diagnosis, the
same algorithm can be used for computing cardinality-minimal
diagnosis. The main difference is that for subset-minimal diagnosis,
$\minhs$ can denote subset-minimal hitting sets, and for
cardinality-minimal diagnosis, $\minhs$ must denote
cardinality-minimal hitting sets.
% % put back in the final version!
% As proposed in earlier
% work~\cite{stuckey-padl05,stern-aaai12,liffiton-cpaior13,pms-aaai13,lpmms-cj16},
% the algorithm iteratively computes minimal diagnoses and minimal
% explanations, and reports one in each main iteration of the
% algorithm. The key objective is to find a new minimal hitting set of
% (all) the explanations, and so a minimal diagnosis, at each iteration
% of the algorithm. If the computed minimal hitting set of the
% explanations is not a diagnosis, then a new (missing) minimal
% explanation is extracted, which is then added to the set of minimal explanations.
%
As proposed in earlier
work~\cite{stuckey-padl05,stern-aaai12,liffiton-cpaior13,lpmms-cj16},
the algorithm iteratively computes minimal diagnoses and minimal
explanations, and reports one in each main iteration of the
algorithm. The key objective is to find a new minimal hitting set of
(all) the explanations, and so a minimal diagnosis, at each iteration
of the algorithm. If the computed minimal hitting set of the
explanations is not a diagnosis, then a new (missing) minimal
explanation is extracted, which is then added to the set of minimal explanations.
If the computed minimal hitting set is indeed a diagnosis (for all
observations), then it is discarded for future iterations by
blocking the same hitting set from being computed.

In contrast with other enumeration approaches proposed
recently~\cite{lpmms-cj16}, which can be viewed as targeting
enumeration of explanations,~\autoref{alg:diag-enum} will terminate as
soon as all the diagnoses have been computed, even if some
explanations have not yet been identified (see
lines~\ref{alg:loop-beg}--\ref{alg:loop-end}). Indeed, as soon as all
diagnoses for some observation have been computed and blocked, one
cannot find another way to recover consistency for that observation.
The lines~\autoref{alg:loop-beg}--\ref{alg:loop-end} can in practice
be made optional if the goal is to compute
some number $K$ of diagnoses.

%% \begin{example}
%%   Let us revisit again the example from~\eqref{eq:cnf:ex01}.
%%   %
%%   If in the first iteration of the {\textbf while}, $t_r$
%%   Consider any diagnosis besides the one that removes the last OR
%%   gates.
%%   %
%%   \autoref{alg:diag-enum} will check each observation against the
%%   computed minimal diagnosis.
%%   %
%%   Clearly, since for the last observation only the removal of the last
%%   OR gate(s) achieves consistency, then any other computed minimal
%%   hitting set will result in another explanation being computed.
%%   %
%%   Thus, for this example, the algorithm will compute a single
%%   diagnosis, as intended.
%% \end{example}

%%As the analysis in~\autoref{sec:motiv} suggests,
%%As the above example illustrates,~
It is important to note that, in theory~\autoref{alg:diag-enum} can
compute an exponentially large number of explanations in between
computed diagnoses.
However, as the experimental results demonstrate, this worst-case
scenario is not observed in practice.

\begin{example}
  Let us consider a system with $r$ failing observations, each of
  which has exactly two explanations: $\{c_1\}$ and $\{c_2\}$.
  Consider some minimal hitting set that does not pick either $c_1$ or
  $c_2$. Then, the next computed explanation will require the missing
  component to be also hit (picked) in future minimal hitting sets.
  %
  % Let us consider a system with $r$ failing observations, each of
  % which is either explained by declaring component $c_1$ abnormal or
  % component $c_2$ abnormal.
  %
  % Consider some minimal hitting set that either does not pick $c_1$ or
  % $c_2$. Then, the computed explanation will require the same
  % component to be hit (picked) in future minimal hitting sets.
  %
  Thus,~\autoref{alg:diag-enum} will compute the correct diagnosis,
  consisting of both $c_1$ and $c_2$ in two iterations.
\end{example}

One essential aspect of the solution proposed
by~\autoref{alg:diag-enum} is that a \emph{single} copy of the system
is used throughout. In the presence of a large number of observations,
this can represent a crucial improvement.

%\paragraph{Discusssion.}
%
%This section shows how hitting set dualization can serve to focus...

%------------------------------------------------------------------------------%

%------------------------------------------------------------------------------%
% File:        res.tex
%
% Description:
%
% Created:     10 Apr 2017.
%
% Author:      Joao Marques-Silva (jpms).
%------------------------------------------------------------------------------%

\section{Preliminary Experimental Results} \label{sec:res}

The experimental evaluation was performed in Ubuntu Linux on an Intel Xeon
E5-2630 2.60GHz processor with 64GByte of memory. The time limit was set to
600s and the memory limit to 10GByte for each individual instance to run. A
prototype of the proposed iterative hitting set dualization (IHSD) approach
referred to as \emph{DEx} (\emph{\textbf{D}iagnosis \textbf{Ex}tractor}) was
implemented in C++ and consists of two interacting parts. One of them computes
subset-minimal or cardinality-minimal hitting sets of the set of explanations.
The other part tests consistency of the system provided that the hitting set
components are disabled.

Enumeration of cardinality-minimal solutions is achieved with the use of an
incremental implementation of the \mxsat algorithm based on soft cardinality
constraints~\cite{schaub-iclp12,mims-jsat15}, which is the state-of-the-art
\mxsat algorithm that won several categories in the \mxsat Evaluation 2015 and
2016. Computing subset-minimal solutions is done with the use of the LBX
algorithm~\cite{mpms-ijcai15} and its further improvements~\cite{mipms-sat16}
for enumerating MCSes for a given unsatisfiable formula. In the performed
evaluation, DEx is configured to compute subset-minimal solutions.
The proposed algorithm was compared to (1) the naive approach of separate
analysis and posterior assemblage and (2) the approach of aggregated analysis.
Both comparisons are detailed below.

\subsubsection{Comparison to separate analysis.} \label{sec:res-sep}
The idea of comparing the proposed approach to separate analysis is to show
that enumerating diagnoses for individual observations can be infeasible in
practice.
Here, we ignore the assemblage phase of the naive approach and focus only on
enumerating diagnoses instead.
However, it should be noted that the assemblage phase would clearly impose an
additional overhead.
As a test material, we constructed a family of \emph{``buggy encoder''} CNF
instances described in \autoref{sec:motiv} (see~\eqref{eq:cnf:ex01}).
For that, we considered the number of observations $r$ varying from $10$ to
$300$ with step $10$, i.e.\ $r\in\{10,20,30,\ldots,290,300\}$.
The number $k$ varies from $2$ to $9$ and also from $10$ to $300$ with step
$10$.
The total number of CNF formulas in the constructed family is 1140.
Recall that the exact number of diagnoses for each of the $r$ observations in
the considered benchmarks is $4^{k} + 4^{\lfloor k/2\rfloor} + 4^{\lceil
k/2\rceil} + 1$, which makes it impractical to enumerate all diagnoses for any
reasonably large $k$.
Here we are aiming at confirming practically that the applicability of the
separate analysis approach is rather limited with respect to the considered set
of instances.
The separate analysis phase of the naive approach is represented in our
evaluation by the two well-known MCS enumerators: RS~\cite{bacchus-aaai14} and
LBX~\cite{mpms-ijcai15}.

% \begin{figure*}[!t]
%   \includegraphics[width=\textwidth]{cactus}
%   \caption{Iterative hitting set dualization vs separate analysis.}
%   \label{fig:cactus}
% \end{figure*}

\begin{figure*}[!t]
  \begin{subfigure}[b]{0.49\textwidth}
    \centering
    \includegraphics[width=\textwidth]{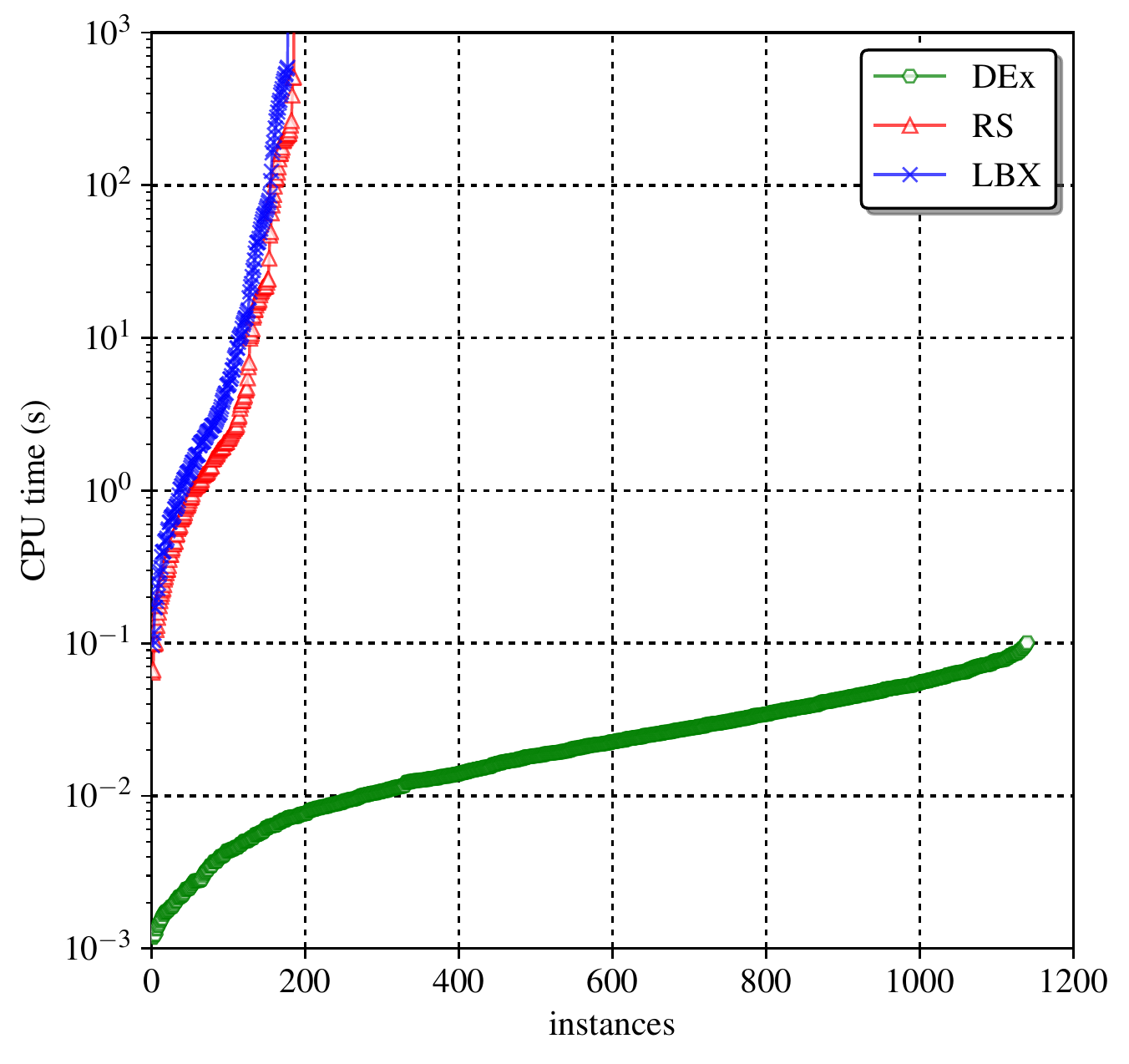}
    \caption{IHSD vs separate analysis}
    \label{fig:cactus}
  \end{subfigure}%
  \begin{subfigure}[b]{0.49\textwidth}
    \centering
    \includegraphics[width=\textwidth]{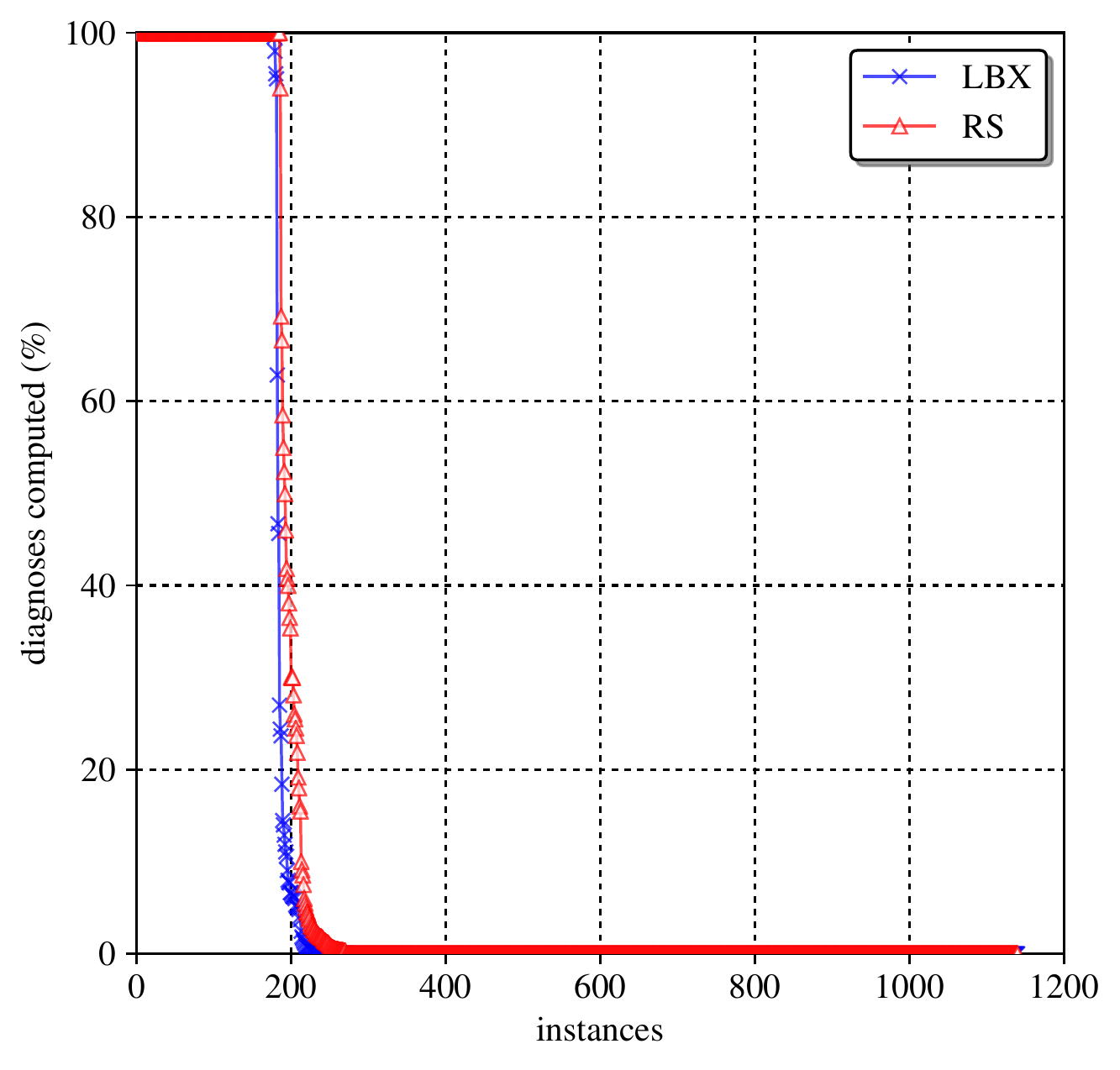}
    \caption{\% of diagnoses computed by LBX and RS}
    \label{fig:perc}
  \end{subfigure}
  \caption{Performance of iterative hitting set dualization and separate analysis.}
  \label{fig:res-sep}
\end{figure*}

\autoref{fig:res-sep} shows the performance of the proposed approach compared
to separate analysis utilizing solvers LBX and RS.
As one can observe in \autoref{fig:cactus}, DEx can efficiently compute the
correct diagnosis for all the 1140 benchmarks.
The time spent for the largest instance, is about $0.1$ second.
In contrast, separate analysis is able to do a comprehensive enumeration of all
diagnoses only for 177 and 185 instances when using LBX and RS, respectively.
Note that both tools can successfully solve only instances with $k<8$.
\autoref{fig:perc} shows how close the LBX and RS solvers are to enumerate all
the diagnoses for each of the individual observations.
As expected, LBX and RS compute 100\% of diagnoses only for the solved
instances, i.e.\ 177 and 185 instances, respectively.
Observe that the percentage of computed diagnoses drops very quickly and is
$\approx0\%$ for most of the instances.

\subsubsection{On aggregated analysis.} \label{sec:res-aggr}
The purpose of the following discussion is to show how dramatically the formula
size grows with the growth of the number of observations $r$ for the example
CNF instance of~\eqref{eq:cnf:ex01}.
\autoref{tab:aggr1} compares the size of the original formula (in terms of the
number of variables and clauses) fed to the proposed IHSD approach and the size
of the formula resulted from aggregating all observations.
Additionally, the table shows the running time of the proposed approach and the
aggregated analysis implemented with the use of LBX and RS.
Note that the formula construction time is excluded here while in practice it
can noticeably contribute to the total running time of the solver.
As one can see, a reasonably large number of observations results in a formula
having a few orders of magnitude more variables and clauses than in the
original formula.
As an example, while the original formula has a few thousand variables and
clauses, the size of the aggregated formula goes beyond millions of variables
and clauses, which makes it hard to build and deal with.
As mentioned earlier, the example of~\eqref{eq:cnf:ex01} is designed to be as
simple as possible while illustrating the main points of the paper.
However, in practice formulas can contain an arbitrarily large number of
additional clauses that do not help revealing the culprits of system's
inconsistency but contribute a lot to the aggregated formula size.
To confirm this, \autoref{tab:aggr2} shows the rate of aggregated formula
growth given the size of some (arbitrary) original formulas.
As one can observe, given a system, which is inconsistent with a few hundred
observations, encoded as a formula with a few hundred of thousands of variables
and a few million of clauses (which may well happen in practice), the
aggregated formula contains millions of variables and clauses, which makes it
hard to deal with and results in the aggregated approach being ineffective in
practice.

\begin{table}[!t]
  \caption{Comparison of the formula size and running time of the IHSD-based
    approach and aggregated analysis for the example of~\eqref{eq:cnf:ex01} for
    growing values of $r$ and $k$.}

  \setlength{\tabcolsep}{0.5em}
  % \small
  \begin{center}
    % \begin{tabular}{cc@{\hskip 0.2in}ccc@{\hskip 0.5in}cccc}
    \begin{tabular}{ccccccccc}
      \toprule
      \multirow{2}{*}{\normalsize $r$} &
      \multirow{2}{*}{\normalsize $k$} &
      \multicolumn{3}{c}{\textbf{\normalsize IHSD approach}} &
      \multicolumn{4}{c}{\textbf{\normalsize Aggregated analysis}} \\
      \cmidrule(lr){3-5}
      \cmidrule(lr){6-9}
      & & variables & clauses & time
      & variables & clauses & LBX & RS \\
      \midrule
      % \cmidrule(lr){3-9}
      10 & 10 &
              50 & 56 & 0.01 &
              592 & 616 & 0.01 & 0.01 \\
      % \midrule
      % \cmidrule(lr){3-9}
      50 & 50 &
              210 & 256 & 0.01 &
              $10\,912$ & $13\,056$ & 0.02 & 0.03 \\
      % \midrule
      % \cmidrule(lr){3-9}
      100 & 100 &
              410 & 506 & 0.02 &
              $41\,812$ & $51\,106$ & 0.08 & 0.09 \\
      % \midrule
      % \cmidrule(lr){3-9}
      200 & 200 &
              810 & $1\,006$ & 0.04 &
              $163\,612$ & $202\,206$ & 0.27 & 0.31 \\
      % \midrule
      % \cmidrule(lr){3-9}
      500 & 500 &
              $2\,010$ & $2\,506$ & 0.19 &
              $1\,009\,012$ & $1\,255\,506$ & 1.75 & 1.84 \\
      % \midrule
      % \cmidrule(lr){3-9}
      500 & $1\,000$ &
              $3\,510$ & $4\,506$ & 0.33 &
              $1\,762\,512$ & $2\,257\,506$ & 3.24 & 3.35 \\
      % \midrule
      % \cmidrule(lr){3-9}
      500 & $2\,000$ &
              $6\,510$ & $8\,506$ & 0.49 &
              $3\,269\,512$ & $4\,261\,506$ & 6.15 & 6.33 \\
      \bottomrule
    \end{tabular}
  \end{center} \label{tab:aggr1}
\end{table}

\begin{table}[!t]
  \caption{Asymptotic growth for the aggregated formulas.}
  \begin{center}
    \setlength{\tabcolsep}{0.5em}
    \begin{tabular}{ccccc}
      \toprule
      \multirow{2}{*}{$r$} &
      \multicolumn{2}{c}{\textbf{Original formula}} &
      \multicolumn{2}{c}{\textbf{Aggregated formula}} \\
      \cmidrule(lr){2-3}
      \cmidrule(lr){4-5}
      &
      variables & clauses &
      variables & clauses \\
      \midrule
      10 &
        100 & 100 &
        $1\,000$ & $1\,000$ \\
      100 &
        $1\,000$ & $1\,000$ &
        $100\,000$ & $100\,000$ \\
      200 &
        $10\,000$ & $10\,000$ &
        $2\,000\,000$ & $2\,000\,000$ \\
      200 &
        $20\,000$ & $100\,000$ &
        $4\,000\,000$ & $20\,000\,000$ \\
      300 &
        $100\,000$ & $1\,000\,000$ &
        $30\,000\,000$ & $300\,000\,000$ \\
      \bottomrule
      \end{tabular}
  \end{center} \label{tab:aggr2}
\end{table}

%------------------------------------------------------------------------------%

%------------------------------------------------------------------------------%
% File:        conc.tex
%
% Description: 
%
% Created:     10 Apr 2017.
%
% Author:      Joao Marques-Silva (jpms).
%------------------------------------------------------------------------------%

\section{Conclusions \& Research Directions} \label{sec:conc}

Model based diagnosis and its many practical instantiations find a
growing number of important practical applications.
Recent work on model based diagnosis has focused on efficiently computing
cardinality-minimal
diagnoses~\cite{codish-aaai12,codish-jair14,msjim-ijcai15} and on
efficiently listing subset-minimal
diagnoses~\cite{stern-aaai12,wotawa-ijcai13}.
Another important problem is how to compute (subset or cardinality)
minimal diagnoses in the presence of multiple (failing) observations,
with the purpose of achieving increased accuracy.
This situation is common in different settings, including software
fault localization and spreadsheet debugging, among others.

This paper shows that existing solutions are bound to be ineffective
in practical domains, for one of two reasons: (i) approaches based on
separate analysis may need to compute a tool large number of
dianogses; and (ii) approaches based on aggregate analysis may need to
analyze too large formulas. As a result, the paper proposes a novel
approach for computing (subset or cardinality) minimal diagnoses in
the presence of multiple (failing) observations, by exploiting
implicit hitting set dualization.
The experimental results confirm both the efficiency and scalability
of the proposed approach.

Future work will validate the performance gains also in the case of
cardinality-minimal diagnoses, which can be implemented seamlessly
using the proposed approach.
%
% Future work will validate the performance gains also in the case of
% cardinality-minimal diagnoses, which can be implemented seamlessly
% using the approach proposed in this paper.
%
In addition, the ideas proposed in this paper will be applied in
different application domains.

%------------------------------------------------------------------------------%

% \clearpage
%\input{proofs}

%------------------------------------------------------------------------------%
% \clearpage  % -> ADD TO FINAL VERSION
\bibliography{refs}
\bibliographystyle{abbrv}
%------------------------------------------------------------------------------%

\end{document}